# A century of oil and gas exploration in Albania: assessment of Naturally Occurring Radioactive Materials (NORMs)

G. Xhixha[1,*], M. Baldoncini[2], I. Callegari[1], T. Colonna[3], F. Hasani[4], F. Mantovani[2], F. Shala[5], V. Strati[2], M. Xhixha Kaçeli[1]

[1] Legnaro National Laboratory, Istituto Nazionale di Fisica Nucleare (INFN), Via dell'Università, 2 - 35020 Legnaro, Padova, Italy

[2] Department of Physics and Earth Sciences, University of Ferrara, Via Saragat, 1 - 44100 Ferrara, Italy

[3] Center for GeoTechnologies, University of Siena, Via Vetri Vecchi, 34 - 52027 San Giovanni Valdarno, Arezzo, Italy

[4] Kosovo Agency for Radiation Protection and Nuclear Safety (KARPNS), Office of the Prime Minister, Ish-Gërmia - 10000 Prishtinë, Kosovo

[5] Faculty of Mechanical Engineering, University of Pristina "Hasan Prishtina", BreguiDiellit – 10000 Pristina, Kosovo

* Corresponding author: Tel. +39 329 6965933; E-mail: xhixha@fe.infn.it (G. Xhixha).

## Abstract

Because potential Naturally Occurring Radioactive Materials (NORMs) generated from oil and gas extractions in Albania have been disposed without regulatory criteria in many decades, an extensive survey in one of the most productive regions (Vlora-Elbasan) has been performed. Among 52 gamma-ray spectrometry measurements of soil, oil-sand, sludge, produced water and crude oil samples, we discover that relatively low activity concentrations of $^{226}$Ra, $^{228}$Ra, $^{228}$Th and $^{40}$K, which are 23 ± 2 Bq/kg, 23 ± 2 Bq/kg, 24 ± 3 Bq/kg and 549 ± 12 Bq/kg, respectively, come from oil-sand produced by hydrocarbon extraction from molasses formations. The mineralogical characterization together with the $^{228}$Ra/$^{40}$K and $^{226}$Ra/$^{40}$K ratios of these Neogene deposits confirm the geological and geodynamic model that predicts a dismantling of Mesozoic source rocks. The average activity concentrations (± standard deviations) of the radium isotopes ($^{226}$Ra, $^{228}$Ra) and of the $^{228}$Th and $^{40}$K radionuclides in soil samples are determined to be 20 ± 5 Bq/kg, 25 ± 10 Bq/kg, 25 ± 9 Bq/kg and 326 ± 83 Bq/kg, respectively. Based on these arguments, the future radiological assessment of other fields in the region can be strategically planned focusing on the oil-sands from molasses sediments. No disequilibrium in the $^{228}$Ra decay segment has been observed in soil, sludge and oil-sand samples within the standard uncertainties. After a detailed radiological characterization of the four main oilfields, we can conclude that the outdoor absorbed dose rate never exceeds the worldwide population weighted average absorbed dose rate in outdoor air from terrestrial gamma radiation.

## 1. Introduction

Petroleum exploration and production in Albania began in the second decade of the 20th century in the Vlora-Elbasan Region, which is characterized by Neogene clastic deposits. The primary oil extraction technique in Albania uses beam pumps, which exploit the pressure of the gas in the reservoir, forcing oil out and into the well. Recently, several secondary recovery techniques, such as water and steam injection, have been employed in pilot wells in the area, leading to an increase in the oil production rate. In the future, the use of unconventional methods of shale gas extraction such as "fracking" is not excluded. All these techniques may cause Naturally Occurring Radioactive Materials (NORMs) to rise to the surface as part of the flow back and production brine. In general in the oil and gas industry, specific attention is dedicated to the contamination of the oil equipment and of the oilfield area by NORMs (e.g., scale, sludge and produced water).

Several environmental studies have shown that the release of produced water into the oil well's surroundings, into decantation plants and into oil spillage sites can result in serious soil, water and air contamination from BTEX (acronym for Benzene, Toluene, Ethylbenzene, and Xylenes), volatile organic compounds or crude oil (REC 2000; Beqiraj et al. 2010; Guri et al. 2013). However, after a century of oil exploration in Albania with poor regulatory criteria, especially in the early beginnings, the risk of soil contamination from NORMs needs to be investigated. According to the CD 2013/59/EURATOM (2014) directive, the identification and monitoring of industrial processes involving NORMs should be conducted by countries to assess the radioactive exposure to workers or members of the public.

NORMs produced by the oil and gas industry are residues enriched with radium isotopes that originate from the uranium and thorium present in reservoir rocks. Indeed, whereas both uranium and thorium are present in hydrocarbon reservoir rocks and are essentially insoluble under reducing conditions, their progenies $^{226}$Ra and $^{228}$Ra concentrate in formation waters. For this reason, $^{226}$Ra and $^{228}$Ra are unsupported by the long-lived uranium and thorium parent radionuclides, and due to their half-lives of 1600 yr and 5.75 yr, respectively, they tend to accumulate in formation water. Produced water (i.e., water brought to the surface) may contain various cations in solution, such as barium, calcium, and strontium, as well as sulfate and carbonate anions, with which radium can consequently co-precipitate as radium sulfates or radium carbonates. This effect can lead to the subsequent formation of scales in oil and gas equipment, for which the $^{226}$Ra and $^{228}$Ra specific activities can be as high as $1500 \times 10^4$ Bq/kg and $280 \times 10^4$ Bq/kg, respectively, though highly variable concentrations are typically observed (Xhixha et al. 2013). Furthermore, sludge wastes (mixture of crude oil, oil-sand and soil) can also be generated, which show slightly lower activity concentrations with respect to scales, up to $350 \times 10^4$

Bq/kg and 205 × 10$^4$ Bq/kg, respectively, for $^{226}$Ra and $^{228}$Ra (Xhixha et al. 2013). However, no study up to now has been performed on oil-sands wastes produced during heavy oil extraction.

This study is the first investigation of the radioactivity concentration in soils, oil-sands, sludges, crude oil and produced water from the KUçova (KU), MArinza (MA), BAllsh (BA) and HEkali (HE) oilfields. This original dataset will provide baseline information concerning possible contaminations due to oil-sands. The geological framework and mineralogical analysis provide an exhaustive interpretation of the radiometric data. This analysis will allow the assessment of the environmental and human impact of the oil industry's activities on the four areas. Moreover, these results may lead to the implementation in the Albanian legislation of the European recommendations regarding the "Basic Safety Standards for the protection against dangers arising from exposure to ionizing radiation (CD 2013/59/EURATOM, 2014)".

## 2. Material and Methods

### 2.1. Geological setting of the study area

Albania is part of the Alpine Mediterranean Mountain belt and can be subdivided into two groups of tectonic units with NNW - SSE orientation: the Inner Units, characterized by intense deformation comprising ophiolites and metamorphic rocks, and the Outer Units, which are made up of regional thrusts involving sedimentary rocks of Triassic to Pliocene age (Prifti and Muska 2013). The main rocks characterizing the units of the two groups, according to the tectonic scheme, are shown in Figure 1. The Outer Units are affected by a major transfer zone, i.e., the Vlora-Elbasan lineament, along which most of the oil and gas fields occur (Roure et al. 2010).

The onshore petroleum extraction in Albania is primarily located in the IONian unit (ION) and in the Durrës Basin, also referred to as the Peri-Adriatic Depression (PAD). The ION is an unbroken, elongate tectonic unit (ca. 60-70 km long x 60 km wide) that extends continuously southwards into Greece. In the ION, two main tectonics phases are recognized (i.e., Middle Miocene and Miocene-Pliocene), forming large anticlines and synclines cut by major high-angle reverse faults (Robertson and Shallo 2000).

The exposure in the ION begins with evaporites and then passes upwards into shallow-marine platform carbonates (limestone and dolomites) of Late Triassic-Early Jurassic age. This succession culminates in a hardground (ancient lithified seafloor) overlain by pelagic carbonates and cherts (Middle–Late Jurassic to Eocene age, having a thickness of 2.5–4 km) and by terrigenous turbidites (Late Eocene-Early Miocene age, with thickness on the order of 4-6.5 km). Paleogene units are disconformably overlain by Neogenic sediments. The PAD is filled by molasses mega-sequences of conglomerates, sandstone and clastic limestones, with clays, shale, coal and gypsum at the top. The thickness of the molasses deposits increases from southeast to northwest, reaching a maximum of 5 km (Silo et al. 2013).

According to **Silo et al. (2013)**, we can recognize six hydrocarbons potential source rocks in the Mesozoic limestone stratigraphic column of the ION. The main ones are the Triassic bituminous dolomite at the bottom, the lower Jurassic schists and other Jurassic source rocks. We also find Cretaceous shales under thrust synclines, which in the future could be exploited as hydrocarbon generators.

The migration of crude oil from the Mesozoic limestone towards the Neogene reservoirs results from successive episodes driven by the aforementioned tectonic structures. The oil currently explored in the Neogenic molasses sediments is the product of a secondary migration from underlying limestone (ION). The hydrocarbons are generally accumulated in stratigraphic traps of the PAD's clastic deposits. Because these phenomena evolved over geological time with different horizontal and vertical distributions, each reservoir displays its own characteristics due to the local geochemical conditions.

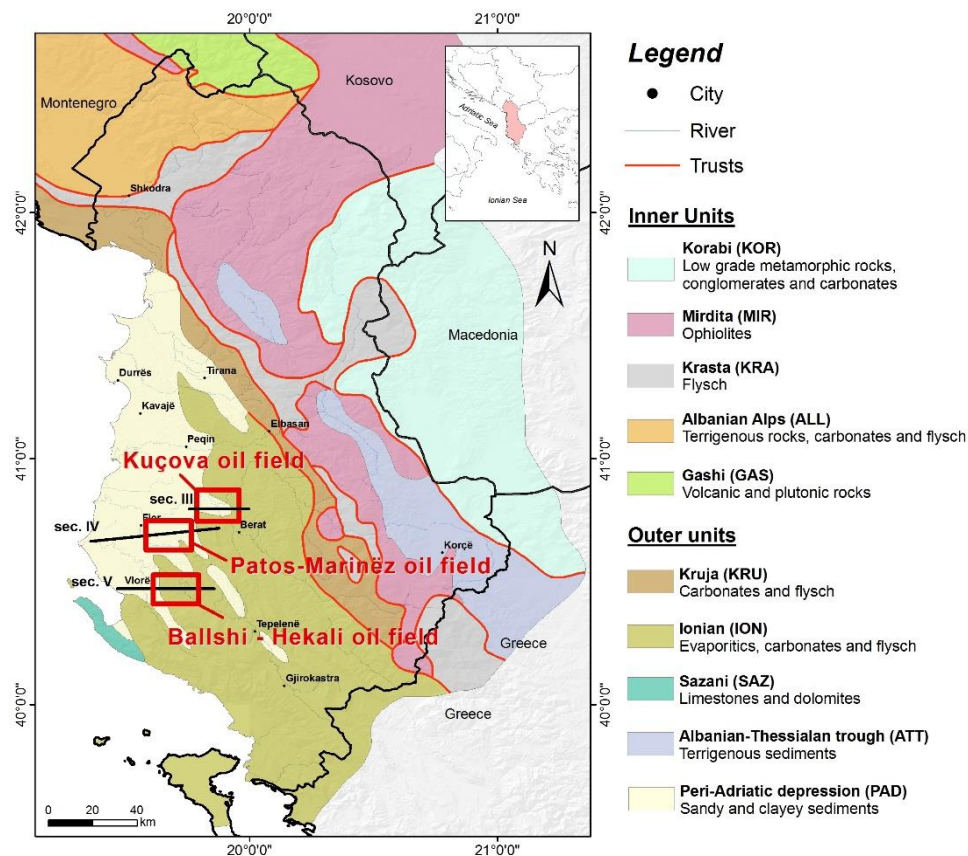

**Figure 1**. Simplified geological map of Albania modified from **Havancsák et al. (2012).** The labels refer to the geological units. The red squares represent the locations of the sampling areas; the solid black lines show the sections (not to scale).

## 2.2. Sample collection

### 2.2.1. Kuçova oilfield

The Kuçova (KU) oilfield is located in the central eastern part of the Durrës Basin and extends along a flat surface of approximately 14 km$^2$. The area is covered by alluvial sediments that are investigated with 21 soil samples (**Fig. 2 (KU)**), collected at a depth of 0–10 cm, with the aim of characterizing the terrestrial component of the outdoor absorbed dose rate. According to the classification of USDA Soil Taxonomy, the textural characteristics of the soil samples are overall sandy (percentage of sand > 50% for most of the samples) with medium amounts of silt (approximately 30%) and low amounts of clay (<20%).

The hydrocarbon reservoir is set in the Driza, Gorani, Kuçova and Polovina members of the Tortonian-Messinian molasses sediments (PAD). In particular, the oil being extracted is accumulated in sand lenses (**Fig. 3 sect. AA'**) (extending from 100 to 1500 m) sealed by shales (**Sejdini et al. 1994**) with a high amount of clay (**Prifti and Muska 2013**). The depth of the reservoir has a range variable from the surface down to 1500 m.

The KU oilfield was discovered in 1928 and Bankers Petroleum Ltd., a Canadian company with the full rights to develop it, declared in 2013 approximately 297 million barrels of Original Oil in Place (OOIP). The oil extracted is characterized by API gravity values and sulfur content belonging to the ranges 14-22° and 4-5%, respectively (**Sejdini et al. 1994**; **Prifti and Muska 2013**). During the periodical pipe cleaning process, we collected 10 samples of oil-sand from three different operative wells (500 – 1000 m depth) and 3 samples of sludge from the surrounding area. Furthermore, produced water and crude oil were collected in polyethylene bottles directly from the decantation plants.

### 2.2.2. Marinza oilfield

The Marinza (MA) oilfield, located in the southeastern part of the Durrës Basin, is the northern part of the Patos-Marinza oilfield system that extends for 5 km WE and 14 km NS. The wide plain is covered by alluvial sediments that are investigated for radiological assessment with 3 soil samples (**Fig. 2 (MA)**) classified according to their texture as clay-loam with sand < 40% (less than Kuçova soil), clay at 35% and silt at 25%.

The Driza, Gorani, and Marinza members of the Tortonian-Messinian molasses formation (PAD) are the main hydrocarbon reservoirs. The multiple layers of unconsolidated fine grain sandstone are located at a depth ranging from 900 to 2000 m. The depositional environment is a marine environment with fluvial-channel and shore face deposits. In some cases, the underlying fractured carbonates (ION), which are located below a major unconformity (**Fig. 3 sect. BB'**), can be oil reservoirs (**Weatherill et al. 2005**; **Sejdini et al. 1994**). The Patos-Marinza oilfield was discovered in 1929 and the amount of OOIP is approximately 5.4 billion barrels, according to 2013 data reported by Bankers Petroleum Ltd.

The two separate fields (Patos in the south and Marinza in the north) have productive sands at different depths, i.e., 0 - 1200 m for Patos and 1200 - 1800 m for MA. The oil extracted is characterized by an API gravity value range of 3 - 33° and a sulfur content range of 2-7% (Weatherill et al. 2005; Prifti and Muska 2013). We collected 1 sludge sample and 1 oil-sand sample for each of the two investigated wells (2000 and 2500 m depth) in the MA oilfield.

### 2.2.3. Ballsh-Hekali oilfield

The Ballsh-Hekali oilfield system is located in a hill area 29 km SE of Fieri. The radiological assessment of the two distinct fields, Ballsh (BA) and Hekali (HE), was performed by collecting 3 and 5 soil samples, respectively (Fig. 2 (BA) and Fig. 2 (HE)), whose textures are classified as sand or loamy sand with sand > 80%, clay < 10% and silt at 20%.

The reservoirs are restricted to Paleocene-Eocene carbonate rocks (ION) and have a depth range of 1000 - 3000 m. The geological structure consists of two superposed faulted anticlines (Fig. 3 sect. C-C'), and the oil stratigraphic column has a thickness greater than 550 m (Sejdini et al. 1994). The BA and the HE oilfields were discovered in 1966 and the OOIP is approximately 440 million barrels (Sejdini et al. 1994). The oil extracted is characterized by an API gravity value of 2-30°, with a sulfur content of 3-7.5% (Prifti and Muska 2013). For the radiological characterization, 2 samples of sludge have been collected in an operative well of the BA oilfield.

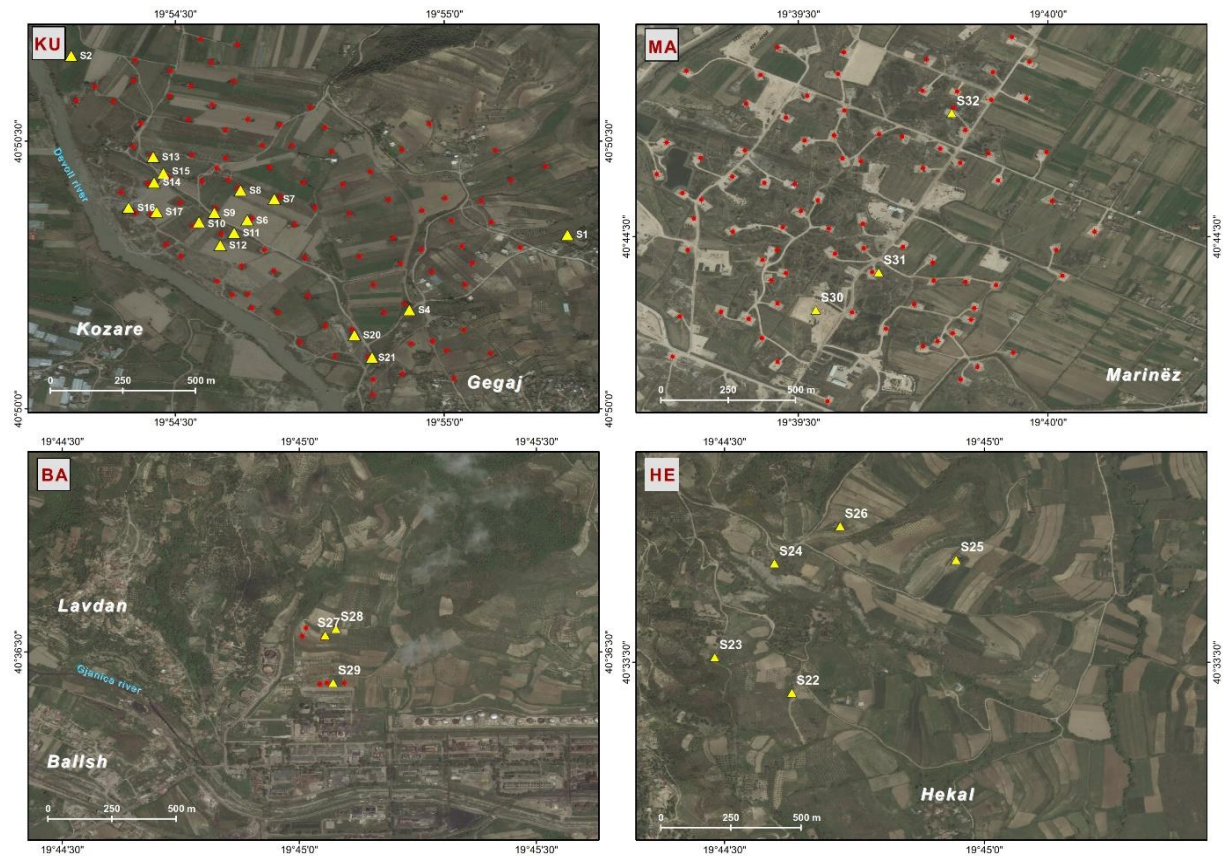

Figure 2. Approximate locations of the oils wells (red stars) and of the soil samples collected in the study area (yellow triangles).

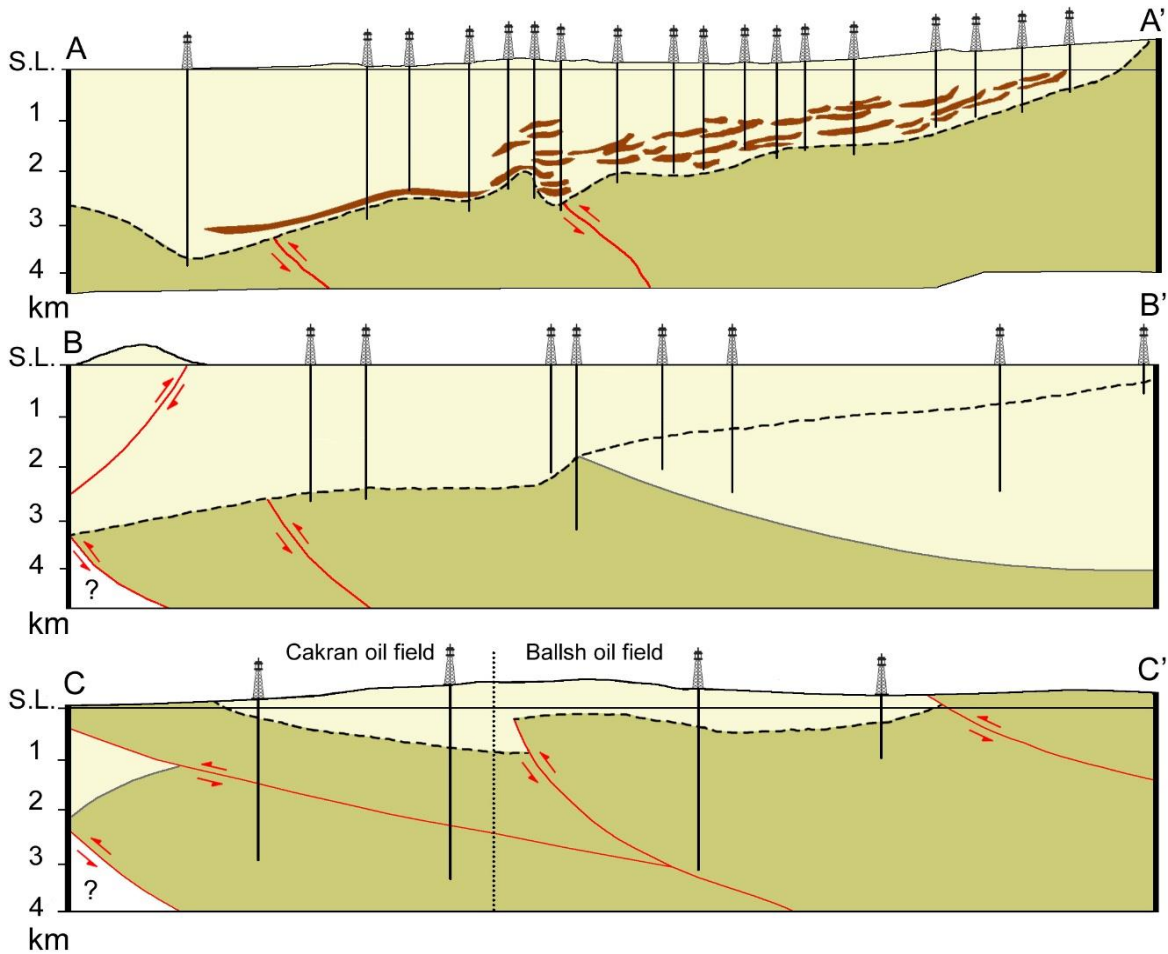

Figure 3. Cross sections a) A-A' in the KU oilfield, b) B-B' in the MA oilfield and c) C-C' in BA-HE oilfield (horizontal dimensions are not to scale), modified from **Silo et al. (2013)**, obtained from the seismic lines shown in **Fig. 1**. The colors of the sections are consistent with that of **Fig. 1** and correspond to the PAD (yellow) and ION (green) zones.

### 2.3. Sample preparation and measurements

### 2.3.1. HPGe gamma-ray spectrometry measurements

The samples were homogenized to a grain size of less than 2 mm and dried for at least 24 h at a temperature of 105°C until a constant weight was achieved. Produced water and crude oil were collected in polyethylene bottles directly from the decantation plants at the KU oilfield. The samples were then transferred in cylindrical polycarbonate boxes of 180 cm$^3$ volume, sealed hermetically and left undisturbed for at least four weeks prior to being measured with a HPGe gamma spectrometer to establish radioactive equilibrium in the $^{226}$Ra decay chain segment.

The samples were measured using the MCA_Rad system described in Xhixha et al. (2013). The fully automated spectrometer consists of two 60% relative efficiency coaxial p-type HPGe gamma-ray detectors with an energy resolution of ~1.9 keV at 1332.5 keV ($^{60}$Co). The absolute full energy peak efficiency of the MCA_Rad is calibrated using certified standard point sources ($^{152}$Eu and $^{56}$Co). The overall uncertainty in the efficiency calibration is estimated to be less than 5%. A certified reference material containing contaminated bulk soil from a Syrian oilfield (IAEA-448) was used for quality control. In general, the $^{226}$Ra, $^{228}$Ac, $^{212}$Pb, $^{208}$Tl and $^{40}$K activity concentrations agreed within 3–10% with respect to the reference values.

The radionuclides studied in this work are $^{226}$Ra, $^{228}$Ra, $^{228}$Th and $^{40}$K by analyzing different gamma lines (Santawamaitre et al. 2014). The presence of $^{137}$Cs in soils was also investigated as a proxy of Chernobyl fallout. The $^{226}$Ra activity concentration was determined by analyzing the two main gamma emissions of the radon progenies, $^{214}$Pb (at 352 keV) and $^{214}$Bi (at 609 keV), and calculating the weighted average. $^{228}$Ra was determined through its direct progeny $^{228}$Ac, giving rise to gamma emissions at 338 keV and 911 keV. The $^{228}$Th activity concentration was determined by analyzing the two main gamma emissions of radon progenies, $^{212}$Pb (at 239 keV) and $^{208}$Tl (at 583 keV). The activity concentrations of $^{40}$K and $^{137}$Cs were determined from their respective gamma emissions at 1460 keV and 662 keV. The acquisition time was set to 4 hours for soil, oil-sand and sludge samples and 24 hours for produced water and crude oil samples.

### 2.3.2. X-ray diffractometry measurements

Before undergoing mineralogical analysis, the samples were dried at 50°C for a few hours and gradually grounding on agate mortar to preserve the crystalline structure of the minerals. The XRD patterns of oriented and random samples were recorded using a GNR APD2000PRO diffractometer, with Cu Kα radiation and a graphite monochromator, operating at 40 mA and 40 kV. X-ray diffractometry of 5 samples was carried out with the divergence and scatter slits set at 1° and the receiving slit at 0.2 mm. The step size and the counting time were 0.03° (2θ) and 3s/step, respectively. A quasi-random orientation of powder samples was obtained by filling a side-entry aluminium holder. The XRD patterns were processed by the SAX Analysis software and the identification of minerals was based on the comparison with PDF-2 reference data supplied from the International Centre for Diffraction Data (ICCD 2013). A semi-quantitative analysis of the minerals identified by XRD was calculated based on chemical analysis, thermogravimetric data related to volatile components and the Reference Intensity Ratio method.

## 3. Results and Discussion

### 3.1. Activity concentrations in soil samples

The results for the $^{40}$K, $^{226}$Ra and $^{228}$Ra and $^{228}$Th activity concentrations with ±1σ uncertainty (in Bq/kg) in soil, oil-sand and sludge samples are shown in **Appendix A**. **Table 1** shows the average activity concentrations at ±1σ standard deviation and the range for soil, sludge and oil-sand samples from different oilfields. The overall averages of the activity concentrations (± standard deviations) of $^{40}$K, $^{226}$Ra, $^{228}$Ra and $^{228}$Th in soil samples are found to be 326 ± 83 Bq/kg, 20 ± 5 Bq/kg, 25 ± 10 Bq/kg and 25 ± 9 Bq/kg, respectively. The results for soil samples are found to be lower or comparable to the global median activity concentrations of $^{40}$K, $^{238}$U and $^{232}$Th, which are 400 Bq/kg, 35 Bq/kg and 30 Bq/kg (**UNSCEAR 2000**), respectively. Moreover, we compared our results with data reported in several studies on the activity concentrations in soil samples from oilfields (**Table 1**). These data show a great variability, which can be attributed to natural features, such as geological and geochemical characteristics of different areas, and to human activities, such as the presence of contamination due to scales and sludges. Another case of study on natural radioactivity in Qatar (**Al-Sulaiti et al. 2012**) relate the high concentration of $^{226}$Ra in soil samples to the NORM associated to oil and gas extraction processes. In our study, although most soil samples were found to be contaminated by crude oil and probably produced water, we do not observe any clear contamination from a radiological point of view. This is probably due to the fact that measured samples of produced water and crude oil show minimum detectable activities (MDA) for all radionuclides, corresponding to 0.4, 1.1, 0.4 and 1.4 Bq/kg, respectively, for $^{26}$Ra, $^{228}$Ra, $^{228}$Th and $^{40}$K.

**Table 1**. Average activity concentrations (±1σ) for $^{40}$K, $^{226}$Ra, $^{228}$Ra and $^{228}$Th in soil, oil-sand and sludge samples for different oilfields studied in Albania are reported together with the average (±1σ), absorbed dose rates (*DR*) and annual effective dose rates (*AEDR*). Average activity concentrations (±1σ) and ranges for $^{40}$K, $^{226}$Ra, $^{228}$Ra and $^{228}$Th in soil samples compared with those from oilfields reported in different studies.

| Sample type | Country (Author) | | No. | $^{40}$K (Bq/kg) | $^{226}$Ra (Bq/kg) | $^{228}$Ra (Bq/kg) | $^{228}$Th (Bq/kg) | DR (nGy/h) | AEDR (mSv/yr) |
|---|---|---|---|---|---|---|---|---|---|
| Soil | Albania | KU | 21 | 297 ± 48 | 17 ± 2 | 19 ± 5 | 20 ± 5 | 32 ± 5 | 0.04 ± 0.01 |
| | | MA | 3 | 360 ± 42 | 19 ± 3 | 33 ± 6 | 29 ± 5 | 44 ± 5 | 0.05 ± 0.01 |
| | | BA | 3 | 253 ± 75 | 21 ± 7 | 29 ± 8 | 29 ± 9 | 38 ± 11 | 0.05 ± 0.01 |
| | | HE | 5 | 472 ± 41 | 30 ± 1 | 42 ± 6 | 40 ± 4 | 58 ± 4 | 0.07 ± 0.01 |
| | Albania (this study) | | 32 | **326 ± 83** (204 – 535) | **20 ± 5** (12 – 32) | **25 ± 10** (11 – 51) | **25 ± 9** (11 – 44) | **38 ± 11** (32 – 58) | **0.05 ± 0.01** (0.04 – 0.07) |
| | Tunisia (Hrichi et al. 2013) | | 1 | **176 ± 1** | **9 ± 0.1** | **11 ± 0.1**[a] | | **18 ± 0.2** | **0.022** |
| | Kuwait (Abdullah et al. 2008) | | 47 | (191 – 296) | (12 – 25) | (9 – 16)[a] | | - | - |
| | Nigeria (Avwiri and Ononugbo 2012) | | 12 | **263 ± 10** (134 – 395) | **30 ± 1** (16 – 52) | **17 ± 1**[a] (10 – 34) | | **35** (23 – 47) | **0.04** (0.03 – 0.06) |

| | | | | | | | | |
|---|---|---|---|---|---|---|---|---|
| | Nigeria (Jibiri and Amakom 2010) | | 9 | **129 ± 70** (MDA – 248) | **76 ± 34** (30 – 122) | **21 ± 6** [a] (12 – 29) | **53 ± 18** (25 – 73) | **0.06 ± 0.02** (0.03 – 0.09) |
| | Canada (Saint-Fort et al. 2007) | | 21 | - | (10 - 10×10$^3$) | (7 – 260) | (MDA – 10) | - | - |
| | Egypt (Shawky et al. 2001) | | 4 | (MDA - 45×10$^3$) | (18×10$^3$ - 438×10$^3$) | (35×10$^3$ - 987×10$^3$) | (MDA – 9×10$^3$) | | |
| Sludge | Albania | - | 3 | 348 ±115 | 19 ± 4 | 22 ± 4 | 23 ± 6 | 36 ± 8 | 0.04 ± 0.01 |
| | | MA | 1 | 314 ± 18 | 18 ± 3 | 22 ± 6 | 25 ± 3 | 35 ± 6 | 0.04 ± 0.01 |
| | | BA | 2 | 175 ± 89 | 20 ± 11 | 21 ± 7 | 13 ± 1 | 29 ± 3 | 0.04 ± 0.01 |
| Oil-sand | Albania | KU | 10 | 549 ± 12 | 23 ± 2 | 23 ± 2 | 24 ± 3 | 47 ± 2 | 0.06 ± 0.01 |
| | | MA | 2 | 366 ± 49 | 12 ± 2 | 14 ± 4 | 12 ± 4 | 29 ± 2 | 0.04 ± 0.01 |

[a] Reported as $^{232}$Th.

Radium isotopes are generally unsupported in formation water, and because a fraction of them can precipitate with oil-sands during oil extraction, an increase in concentration and disequilibrium in the decay chain can occur. The disequilibrium in the $^{228}$Ra decay segment was studied in terms of the ratio between the activity concentrations of $^{228}$Th and $^{228}$Ra (**Appendix B**). We observed ratio values systematically greater than unity; however, this cannot be considered a strong evidence of disequilibrium within the standard uncertainties. The ratios between the activity concentrations of $^{228}$Ra/$^{228}$Th for soil, sludge and oil-sand samples (the last will be discussed below) and the corresponding goodness of the fit, expressed in terms of reduced $\chi^2$ reported in parentheses, are 1.02 ± 0.07 ($\chi^2 = 0.5$), 0.99 ± 0.20 ($\chi^2 = 3.9$) and 1.05 ± 0.09 ($\chi^2 = 1.6$), respectively. We also observe in this case ratio values close to unity for both sludge and oil-sand samples. However, in the context of the environmental legacy issue, only $^{226}$Ra is of long-term concern because after 25 years only approximately 5% of $^{228}$Ra remains, but the evaluation of secular equilibrium conditions is beyond the capacity of the MCA_Rad system.

The highest concentrations of $^{40}$K, $^{226}$Ra, $^{228}$Ra and $^{228}$Th are systematically observed in the HE oilfield, while the lowest ones in the KU oilfield. The KU and MA oilfields belong to the same hydrographic basin, located in the upstream and downstream sectors of the basin, respectively. This is confirmed by the presence of round pebbles with dimensions on the order of centimeters in the KU oilfield which are not present in fine grain soils of the MA oilfield. The higher concentration of $^{228}$Ra and $^{228}$Th observed in the sediments of the MA oilfield with respect to that of the KU oilfield may be correlated to a selective mechanical deposition, which could be observed in placer deposits. Indeed, as reported by **Ostrosi et al. (1998)**, heavy mineral placers are exploited along the Adriatic Sea's shore. However, the higher activity concentrations observed in the HE oilfield can be a consequence of either different host rocks or differences in the geomorphological structure (slope debris) with respect to the KU and MA oilfields. Indeed, the genetic processes of soils originated on carbonate substrates tend to enrich the sediments, in particular with uranium and thorium; this may be the case for the soils from the HE oilfield (**Greeman et al. 1999**).

$^{137}$Cs was observed only in the soil samples and with highly variable concentrations: the average concentration is 6 ± 4 Bq/kg in the KU oilfield, 14 ± 6 Bq/kg in the BA oilfield, 9 ± 9 Bq/kg in the HE oilfield and 15 ± 7 Bq/kg in the MA oilfield. Indeed, for undisturbed soils with a high presence of minerals, the vertical migration of $^{137}$Cs is rather slow, and a greater fraction of activity concentration is expected in the topsoil (0–10 cm) (UNSCEAR, 2008). The non-presence of $^{137}$Cs in oil-sands can be used as a tracer to discriminate them from soil samples.

The activity concentration ratios (±1σ uncertainty) for $^{228}$Ra/$^{226}$Ra, $^{226}$Ra/$^{40}$K and $^{228}$Ra/$^{40}$K (Fig. 4) (hereafter, assuming secular equilibrium, reported as Th/U, U/K and Th/K, respectively) are summarized in Appendix B and studied to understand the hidden geochemical patterns of the analyzed samples. We observe in general that the activity concentration ratios in the soil samples show a specific signature (confirmed at ±1σ) over the oilfield reservoirs, which allows us to attribute a clastic sedimentary origin to them (main constituents of the PAD unit, see Figure 1). Indeed, the measured activity concentration ratios Th/U, U/K and Th/K are comparable with the average activity concentration ratios in sandstone, which are 1.06, 0.062 and 0.068, respectively, and in shale, which are 1.07, 0.055 and 0.058, respectively (Van Schmus et al., 1995). This is reasonable as the soils originated in alluvial deposits overlaying the PAD unit.

The oil-sand samples (discussed later) also have similar ratios (Fig. 4); collectively interpreting the results from the oil-sands is relatively complex due to their different lithological origins. In the case of the MA oilfield, the activity concentration ratios (Appendix B) indicate an increase in Th and K with respect to an average carbonate rock, characterized by Th/U, U/K and Th/K values equal to 0.25, 0.32 and 0.082, respectively (Van Schmus et al., 1995). This can be possibly ascribed to the lithological characteristics of the MA reservoir rock, which has a significant amount of terrigenous minerals (ION unit described in Figure 1). In the case of the KU oilfield, the activity concentration ratios are particularly similar to sandstone rocks (Van Schmus et al., 1995); the slight difference can be attributed to the highly heterogeneous composition of the sandstone rocks.

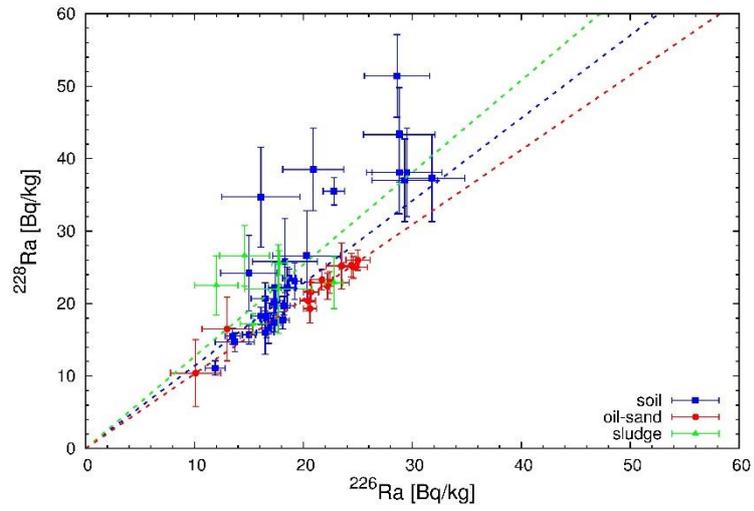

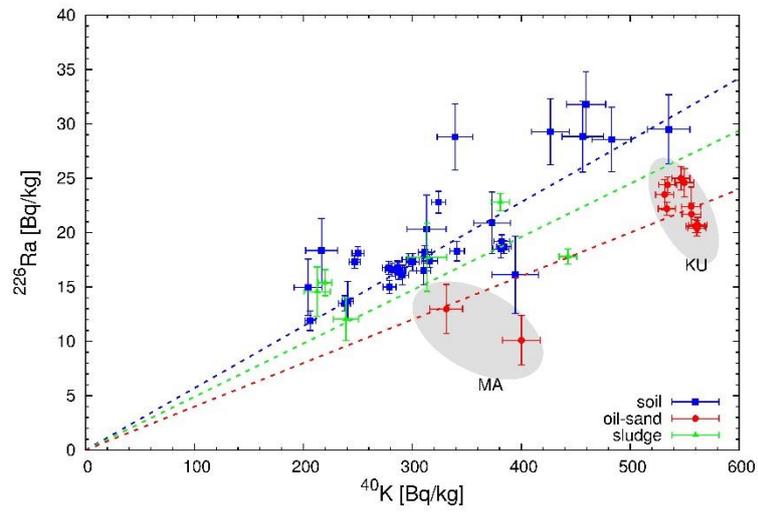

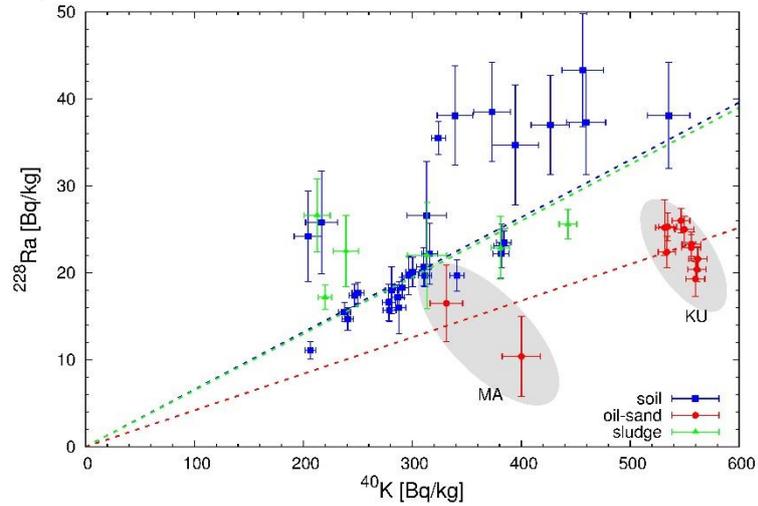

**Figure 4**. (A) $^{228}$Ra/$^{226}$Ra, (B) $^{226}$Ra/$^{40}$K and (C) $^{228}$Ra/$^{40}$K activity concentration ratios for soil (blue), oil-sand (red) and sludge (green) samples.

### 3.2. Activity concentrations in the sludge samples

The averages of the activity concentrations of $^{40}$K, $^{226}$Ra, $^{228}$Ra and $^{228}$Th in the sludges are found to be comparable, within the uncertainty, for all the oilfields investigated (Table 1). As the sludges are a mixture of crude oil, oil-sand and soil, the variability observed in their activity concentrations is difficult to understand as it depends on the mixing fractions. However, we can state that a significant fraction of crude oil will result in a decreasing radioactivity level because the concentrations of radionuclides are very low. In the case of sludge samples from the KU oilfield, we observe a slight increase in the activity concentrations of $^{40}$K, $^{226}$Ra, $^{228}$Ra and $^{228}$Th with respect to the soil samples; this is probably due to the relatively high concentrations of oil-sand in the sludge samples. This assumption is confirmed in the case of the MA oilfield, where we observe a slight decrease in the activity concentrations of $^{40}$K, $^{226}$Ra, $^{228}$Ra and $^{228}$Th with respect to the soil samples due to the relatively low concentrations of oil-sand in the sludge samples.

### 3.3. Activity concentration in the oil-sand samples

The activity concentrations of $^{40}$K, $^{226}$Ra, $^{228}$Ra and $^{228}$Th in the oil-sand samples from the KU and MA oilfields differ significantly (within 2σ uncertainty) (Table 1). The two oilfields can be easily recognized using the patterns of the $^{226}$Ra/$^{40}$K and $^{228}$Ra/$^{40}$K ratios (Figure 4b and 4c). The discrimination between the two oilfields has also been achieved via a mineralogical analysis (Table 2) performed for oil-sand samples collected in different wells from the KU and MA oilfields. These results confirm that the KU oilfield reservoir is located in sandy lenses belonging to the PAD (absence of dolomite and very low calcite, along with the presence of muscovite and chlorite and an increase in K-feldspar), while the MA oilfield reservoir is located in the fractured limestones of the ION (presence of calcite and dolomite) (see Figure 1). This interpretation is also supported by the observation of a higher radioactivity content in the KU oilfield samples (more radioactive sandy lenses) with respect to to that of the MA oilfield (less radioactive limestones) (Figure 4).

**Table 2**. Mineral percentage content measured by X-ray diffractometry and activity concentration measured by the MCA_Rad system for oil-sand samples collected in the KU and MA oilfield. The uncertainty associated to the mineralogical analysis is approximately 10-15% for Quartz and 20-30% for the other minerals (minerals estimated with an uncertainty greater than 30% are reported as <5). The Limit of Detection (LD) for XRD measurements is 1%.

| Sample ID | Qtz | Pl | Kfs | Cal | Dol | Ms | Chl | $^{40}$K (Bq/kg) | $^{226}$Ra (Bq/kg) | $^{228}$Ra (Bq/kg) | $^{228}$Th (Bq/kg) |
|---|---|---|---|---|---|---|---|---|---|---|---|
| KU_OS1 | 40 | 20 | 6 | <5 | <LD | 15 | 7 | 534 ± 8 | 22 ± 1 | 22 ± 2 | 25 ± 2 |
| KU_OS5 | 44 | 20 | 13 | <5 | <LD | 8 | 8 | 547 ± 8 | 25 ± 1 | 26 ± 1 | 29 ± 2 |

| | | | | | | | | | | |
|---|---|---|---|---|---|---|---|---|---|---|
| **KU_OS6** | 43 | 20 | 8 | <5 | <LD | 10 | 6 | 560 ± 9 | 21 ± 1 | 19 ± 2 | 21 ± 1 |
| **MA_OS1** | 60 | 12 | 8 | 12 | <5 | <LD | <5 | 331 ± 15 | 13 ± 2 | 16 ± 4 | 9 ± 2 |
| **MA_OS2** | 55 | 11 | 8 | 20 | <5 | <5 | <LD | 400 ± 18 | 10 ± 2 | 10 ± 5 | 14 ± 2 |

*Qtz.* Quartz, *Pl.* Plagioclase, *Kfs.*K-Feldspar, *Cal.*Calcite, *Dol.* Dolomite, *Ms.* Muscovite, *Chl.*Chlorite

The KU oil-sand samples are characterized by a significant terrigenous content (deep-sea sediments transported to the oceans by rivers from land sources) made up of quartz, feldspars, mica, chlorite and a very low amount of calcite. The presence of chlorite and muscovite minerals confirms the geodynamic model that predicts a dismantling of the magmatic rocks belonging to the Inner Units. In particular, the chlorite could be formed by the alteration of other silicate minerals that contain Mg and Fe (e.g., olivine, augite (pyroxene), hornblende, and biotite), which is typical of ophiolitic suites (deep-sea marine sediments overlying, from top to bottom, pillow basalts, sheeted dikes, gabbro, dunite, and peridotite). The percentages of Quartz, Feldspar and Lithics (QFL) (plotted in the ternary diagram of **Figure 5**) measured in 3 samples of KU oil-sand are comparable with those reported in **Dickinson (1985)**, which correspond to mineralogical signatures typical of terrigenous Cenozoic sandstones deposited in different sites. Referring to the QFL triangle of **Dickinson (1985)**, we can affirm that the oil-sand samples can be related to a source rock environment, such as a "dissected arc", which is connected to the exposure of volcanic and basement rocks (in this case the ophiolitic rocks of the Inner unit). Because the reservoir is formed by sands containing many minerals of magmatic derivation (muscovite, chlorite and feldspars), this may confirm the higher amount of radionuclides with respect to MA.

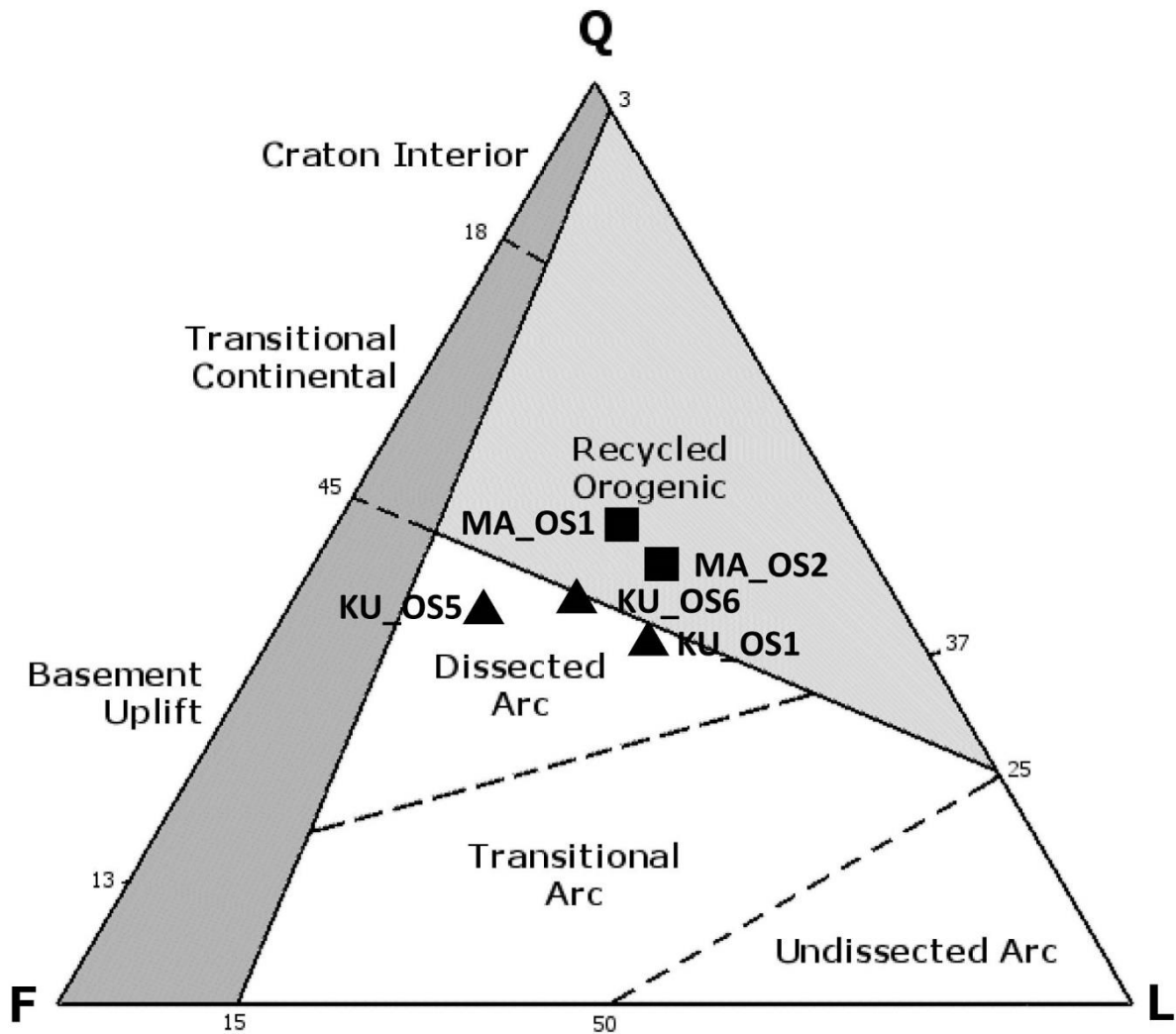

**Figure 5**. Quartz, Feldspar and Lithics ternary diagram after **Dickinson (1985)**. The studied oil-sand plot mainly in the dissected arc and recycled orogenic fields for KU (triangles) and MA (rectangles) oilfields.

In the MA oilfield, the depth provenance of oil-sand is greater with respect to that of the KU oil-sand samples due to the geological structure and to the existence of deep stratigraphic traps. In particular, the MA oil-sand samples come from a reservoir that is located near a planar discontinuity called a "major unconformity" (dashed line in **Fig. 3 Sect. BB'**). The higher calcite content (**Table 2**) of both MA oil-sand samples indicates that the reservoir is located in the carbonate source rock (ION) or near the interface with overlying molasses sediments (PAD) (see **Figure 1**). The data from MA and KU in the QFL (Quartz, Feldspar and Lithics) confirm our analysis. In particular, the MA oil-sands are the result of recycled orogens. The deformation and uplift of supracrustal strata include sedimentary and volcanic rocks exposed in varied fold-thrust belts of orogenic regions. The lower radioactivity content fits perfectly with the geodynamic framework and the mineralogical nature of the oil-sand reservoir. In particular, the source rock with clear calcite content, which is typical for the ION, seems to lack radionuclides, making the extracted oil-sands a negligible source of NORM.

## 4. Radiological Assessment

The absorbed dose rate (*DR*) in air from external gamma radiation at 1 m above ground level due to the presence of uniformly distributed natural radionuclides in measured soils is calculated according to **UNSCEAR (2000)**:

$$DR \text{ (nGy/h)} = 0.0417 A_K + 0.462 A_U + 0.604 A_{Th} \qquad (1)$$

where $A_K$, $A_U$, and $A_{Th}$ are the activity concentrations (in Bq/kg) for $^{40}$K, $^{238}$U (as $^{226}$Ra) and $^{232}$Th (as $^{228}$Ra), respectively. Secular equilibrium was assumed for the dose calculation. The average outdoor absorbed dose rate (at ±1σ uncertainty) in the KU oilfield area is 32 ± 5 nGy/h; in MA 44 ± 5 nGy/h; in BA 38 ± 11 nGy/h; and in HE 58 ± 4 nGy/h (**Table 1**). The calculated outdoor absorbed dose rates range from 21 (KU) to 64 nGy/h (HE). This dose rate is lower than or comparable to the population weighted average absorbed dose rate in outdoor air from terrestrial gamma radiation (60 nGy/h) (**UNSCEAR 2000**).

The average outdoor absorbed dose rates (at ±1σ uncertainty) from the sludge and oil-sand samples are reported in **Table 1**, which range between 27 and 50 nGy/h. Assuming the worst case scenario in the KU oilfield, which corresponds to oil-sands uniformly distributed over the topsoil, the absorbed dose rate was estimated to increase on average 50%, i.e., to a maximum value of 47 ± 2 nGy/h. In the same hypothesis, we observe a decrease in the absorbed dose rate for the other oilfields. Therefore, negligible radiation exposure to workers in the oil and gas industry can occur from increased gamma dose rates due to the relatively higher radioactivity content in the oil-sand residues in the KU oilfield.

The radiological hazard for workers and populations living in the oilfield areas is evaluated in terms of the Annual Effective Dose Rate (*AEDR*). The evaluation of the *AEDR* was performed by adopting an outdoor time occupancy factor equal to 20% and a conversion factor of 0.7 (Sv/Gy), which accounts for the dose rate's biological effectiveness in causing damage to human tissues.

$$AEDR \text{ (mSv/yr)} = DR \times 10^{-6} \text{ (mGy/h)} \times 8760 \text{ (h/yr)} \times 0.7 \text{ (Sv/Gy)} \times 0.2 \qquad (2)$$

The results concerning the radiological assessment are summarized in **Table 1**. The annual outdoor effective dose rate of 0.05 ± 0.01 mSv/y associated to the radioactive content in soils is lower than the worldwide annual effective dose value of 0.07 mSv/y, as reported by **UNSCEAR (2000)**. Considering the worst-case scenario described above, we predict an increase in the *AEDR* of approximately 0.02 ± 0.01 mSv/y for the KU oilfield. This value is negligible considering the recommended excess limit of an effective dose for the local population (1 mSv/y). Therefore, in the KU, MA, BA and HE oilfields, there is no concern from a radiological point of view.

## 5. Conclusions

This study is the first screening campaign on the identification and monitoring of oil and gas industry processes involving NORMs. Accordingly, the KU, MA and BA-HE oilfields are chosen as pilot study areas, where secondary recovery techniques for oil extraction have recently been introduced. In particular, soil (N = 32), sludge (N = 6), oil-sand (N = 12), produced water (N =1) and crude oil (N = 1) samples are measured using the gamma-ray spectrometry method. The results reveal baseline information for local environmental legacy policies and the implementation of EU legislations.

The average activity concentrations (± standard deviations) of $^{40}$K, $^{226}$Ra, $^{228}$Ra and $^{228}$Th in the soil samples collected in the main oilfields in Albania are 326 ± 83 Bq/kg, 20 ± 5 Bq/kg, 25 ± 10 Bq/kg, 25 ± 9 Bq/kg and 6 ± 4 Bq/kg, respectively. The results for the soil samples are found to be comparable with data reported in several studies on activity concentrations in soil samples from oilfields. The activity concentrations in the soil samples show great variability, which can be attributed mainly to the geological and geochemical characteristics of the different areas and in a lower grade probably to human activities, such as the release of oil-sands and sludges. The impact caused by the contamination of the investigated soils collected in the surroundings of Albanian oilfields by crude oil and produced water is negligible from a radiological point of view, as the activity concentrations measured in the produced water and crude oil samples are found to be very low (below MDA). Geological and geochemical arguments are used to describe the observed activity concentrations in different oilfields. The higher radioactivity content observed in the soils from the HE oilfield with respect to that from the KU oilfield was attributed to both the carbonate origin of the host rocks and to different geomorphological structure. Indeed, bibliographic studies confirm that soils that originate on carbonate substrates tend to enrich the sediments in uranium and thorium. In the case of the KU and MA oilfields, which are part of the same hydrographic basin, the increasing trend for the activity concentrations of thorium progenies gives an indication on the selective deposition and, as supported by bibliographic studies, on the occurrence of heavy placer deposits along the Adriatic Sea's shore. According to the Th/U, U/K and Th/K ratios, the origin of the soil samples can be reasonably connected to alluvial deposits overlaying the PAD unit.

The activity concentrations in the sludge samples are reasonably understood when considering sludges as a mixture of soil, crude oil, and oil-sand. This study is the first one reporting results on the activity concentrations in oil-sand samples generated during heavy oil extraction. The average activity concentrations of the oil-sand samples measured in the KU and MA oilfields differ at 2σ level, and are equal to 549 ± 12 Bq/kg ($^{40}$K), 23 ± 2 Bq/kg ($^{226}$Ra), 23 ± 2 Bq/kg ($^{228}$Ra) and 24 ± 3 Bq/kg ($^{228}$Th) in the KU oilfield and 366 ± 49 Bq/kg ($^{40}$K), 12 ± 2 Bq/kg ($^{226}$Ra), 14 ± 4 Bq/kg ($^{228}$Ra) and 12 ± 4 Bq/kg ($^{228}$Th) in the MA oilfield. Based on the mineralogical analysis of the oil-sand samples, we confirm bibliographic geophysical evidence regarding the origin of the oilfield reservoirs: sandy lenses located in the PAD unit for the KU oilfield and fractured limestone located in the ION unit for the MA oilfield. Indeed, the KU oil-sand samples are characterized by terrigenous contents with very low amounts of calcite, and the presence of chlorite and muscovite minerals confirms the geodynamic model that predicts a dismantling of magmatic rocks in the Inner Unit. On the other side, the MA oil-sands confirm their connection to the geological structure due to the higher amount of calcite. This different origin is also confirmed from the activity concentration results, as the KU samples containing many minerals from magmatic deviation show higher activity concentrations of $^{40}$K, $^{226}$Ra, $^{228}$Ra and $^{228}$Th with respect to the MA samples, which originate from carbonate rocks known to have a lower radioactive content. Based on this geological framework, the future radiological assessment of other fields in the region can be strategically planned to focus on the investigation of the oil-sands from Neogene molasses deposits.

The absorbed dose rate and annual effective dose rate were calculated to assess the radioactive exposure of workers or members of the public due to NORMs produced in the oil and gas industry, in compliance with EU recommendation. Assuming the worst case scenario, i.e., that sludge and oil-sands are uniformly distributed over the topsoil, the absorbed dose rate and the *AEDR* of the KU oilfield were estimated to increase on average by a maximum of 50%, i.e., to a maximum value of 47 ± 2 nGy/h or the corresponding 0.06 ± 0.01 mSv/y. In the other investigated oilfields, the impact of industrial processes has been estimated to be negligible. Considering the worldwide annual effective dose value of 0.07 mSv/y associated to the radioactive content in soils and the recommended excess limit of an effective dose for the local population of 1 mSv/y, we conclude that there is no concern from a radiological point of view in the KU, MA, BA and HE oilfields.

**Acknowledgements**

This work is partly supported by the Fondazione Cassa di Risparmio di Padova e Rovigo, by the Istituto Nazionale di Fisica Nucleare (INFN) by the MIUR (ITALRAD Project) and by NORM4BUILDING COST TU1301 project. The authors are grateful to Giampietro Bezzon, Giampaolo Buso, Giovanni Fiorentini, Gazmira Gjeta, Mariola Goga, Liliana Mou, Carlos Rossi Alvarez, Alessandro Zanon and Enrico Guastaldi for the valuable discussion on the manuscript and Francesco Dellisanti (ANALITICA S.a.s) for the valuable comments on the mineralogical analysis and interpretation and to Fredi Gjika, Gjergji Hatellari and Nuri Hatellari for the technical support during sample collection.

**References**

Abdullah, F.H., Saad, H.R., Farhan, A.R., Sharma, M.M., 2008. Investigation of Naturally Occurring Radioactive Materials (NORM) in Oil Fields and Oil Lakes in Kuwait. Society of Petroleum Engineers. SPE International Conference on Health, Safety, and Environment in Oil and Gas Exploration and Production, 1-6. Doi: 10.2118/111562-MS

Al-Sulaiti, H., Nasir, T., Al Mugren, K.S., Alkhomashi, N., Al-Dahan, N., Al-Dosari, M., Bradley, D.A., Bukhari, S., Matthews, M., Regan, P.H., Santawamaitre, T., Malain, D., Habib, A., 2012. Determination of the natural radioactivity levels in north west of Dukhan, Qatar using high-resolution gamma-ray spectrometry. Applied Radiation and Isotopes, 70, 1344-1350. Doi: 10.1016/j.apradiso.2011.11.015

Avwiri, G.O., Ononugbo, C.P., 2012. Natural Radioactivity Levels in Surface Soil of Ogba/Egbema/Ndoni Oil and Gas Fields.Energy Science and Technology. 4(2), 92-101. Doi: 10.3968/j.est.1923847920120402.427

Beqiraj, I., Drushku, S., Seiti, B., Topi, D., Muçaj, A., 2010. Environmental Problems in Albanian Fields of Production and Processing of the Petroleum.NaturaMontenegrina. 9(3), 673-685.

CD (Council Directive) 2013/59/Euratom of 5 Dec. 2013 laying down basic safety standards for protection against the dangers arising from exposure to ionising radiation, and repealing Directives 89/618/Euratom, 90/641/Euratom, 96/29/Euratom, 97/43/Euratom and 2003/122/Euratom. L13, Vol. 57, ISSN 1977-0677. Doi:10.3000/19770677.L_2014.013.eng

Dickinson, W.R., 1985. Interpreting provenance relations from detrital modes of sandstones.In Zuffa, G.G. (Ed.), Provenance of Arenites. NATO ASI Series C: Mathematical and Physical Sciences Vol. 148, Springer Science – Bussiness Media, B.V., Dordrecht, Holland, pp. 333-361. Doi: 10.1007/978-94-017-2809-6

Greeman, D.J., Rose, A.W., Washington, J.W., Dobos, R.R., Ciolkosz, E.J., 1999. Geochemistry of radium in soils of the Eastern United States.Applied Geochemistry. 14, 365-385. Doi: 10.1016/S0883-2927(98)00059-6

Guri, A., Guri, S., Aliu, A., Lubonja, O., 2013. The Impact of Oil Development Activities on Environment and Sustainable Development in Fier Area.Academic Journal of Interdisciplinary Studies. 2(9), 626-634. Doi: 10.5901/ajis.2013.v2n9p626

Havancsák, I., Koller, F., Kodolányi, J., Szabó, C., Hoeck, V., Onuzi, K., 2012. Chromite-hosted Silicate Melt Inclusions from Basalts in the Stravaj Complex, Southern MirditaOphiolite Belt (Albania). Turkish J. Earth Sci. 21, 79–96. Doi: 10.3906/yer-1010-40

Hrichi, H., Baccouche, S., Belgaied, J-E., 2013. Evaluation of radiological impacts of tenorm in the Tunisian petroleum industry.*Journal of Environmental Radioactivity*. 115, 107-113. Doi: 10.1016/j.jenvrad.2012.07.012

IAEA (International Atomic Energy Agency), 2013.*Certified Reference Material, IAEA-448Radium-226 in soil from oil field*, RS_IAEA-448_Rev.2, Vienna, Austria.

International Centre for Diffraction Data (ICDD), 2013. Powder Diffraction File (PDF), PDF-2 Release.

Jibiri, N.N., Amakom, C.M., 2011. Radiological Assessment of Radionuclide Contents in Soil Waste Streams from an Oil Production Well of a Petroleum Development Company in Warri, Niger Delta, Nigeria. *Indoor and Built Environment*. 20(2), 246-252. Doi: 10.1177/1420326X10378806

Ostrosi, B., Qirici, V., Grazhdani, A., 1998. The heavy minerals shore placers of Adriatic Sea in Albania. Bulletin of the Geological Society of Greece. XXXII(3), 173-177.

Prifti, I., Muska, K., 2013. Hydrocarbon occurrences and petroleum geochemistry of Albanian oils. Ital. J. Geosci. (Boll. Soc. Geol. It.). 132(2), 228-235. Doi: 10.3301/IJG.2012.29

REC (The Regional Environmental Center), 2000. *Country Report Albania*: In *Strategic Environmental Analysis of Albania, Bosnia & Herzegovina, Kosovo and Macedonia*, Albania.

Robertson, A., Shallo, M., 2000.Mesozoic–Tertiary tectonic evolution of Albania in its regional Eastern Mediterranean context.Tectonophysics. 316, 197–254. Doi: 10.1016/S0040-1951(99)00262-0

Roure, F., Andriessen, P., Callot, J.P., Faure, J.L., Ferket, H., Gonzales, E., Guilhaumou, N., Lacombe, O., Malandain, J., Sassi, W., Schneider, F., Swennen, R., Vilasi, N., 2010. The use of palaeo-thermo-barometers and coupled thermal, fluid flow and pore-fluid pressure modelling for hydrocarbon and reservoir prediction in fold and thrust belts. *Geological Society, London, Special Publications*, *348*(1), 87-114. Doi: 10.1144/SP348.6

Saint-Fort, R., Alboiu, M., Hettiaratchi, P., 2007. Evaluation of TENORMs field measurement with actual activity concentration in contaminated soil matrices.*Journal of Environmental Science and Health Part A*. 42, 1649-1654. Doi: 10.1080/10934520701518158

Santawamaitre, T., Malain, D., Al-Sulaiti, H.A., Bradley, D.A., Matthews, M.C., Regan, P.H., 2014. Determination of $^{238}$U, $^{232}$Th and $^{40}$K activity concentrations in riverbank soil along the Chao Phraya river basin in Thailand. Journal of Environmental Radioactivity, 138, 80-86. Doi: 10.1016/j.jenvrad.2014.07.017

Sejdini, B., Costantinescu, P., Piperi, T., 1994. Petroleum Exploration in Albania. In. Popescu, B.M. (Ed.), Hydrocarbons of Eastern Central Europe: Habitat, Exploration and Production History. Springer-Verlag Berlin Heidelberg, pp. 1-27.

Shawky, S., Amer, H., Nada, A.A., Abd El-Maksoud, T.M., Ibrahiem, N.M., 2001.Characteristics of NORM in the oil industryfrom Eastern and Western deserts of Egypt.*Applied Radiation and Isotopes*. 55, 135-139. Doi: 10.1016/S0969-8043(00)00364-X

Silo, V., Muska, K., Silo, E., 2013. Hydrocarbon evaluation aspects in Neogene clastic reservoirs, Vlora-Elbasan Region, Albania. *Ital. J. Geosci. (Boll. Soc. Geol. It.)*. 132(2), 220-227. Doi: 10.3301/IJG.2013.04

UNSCEAR (United Nations Scientific Committee on the Effects of Atomic Radiation), 2000.*Exposures from Natural Radiation Sources*. UN, New York.

UNSCEAR (United Nations Scientific Committee on the Effects of Atomic Radiation), 2008. *Health effects due to radiation from the Chernobyl accident*. UN, New York.

Van Schmus, W. R. (1995). Natural Radioactivity of the crust and mantle. In Ahrens, T.J. (Ed.) Global Earth Physics, A Handbook of Physical Constants AGU Ref. Shelf, 1 AGU, Washington, D.C., pp. 283–291.

Weatherill, B., Seto, A.C., Gupta, S.K., Çobo, L., 2005. Cold Heavy Oil Production at Patos-Marinza, Albania. SPE International Thermal Operations and Heavy Oil Symposium, 1-18. Doi: 10.2118/97992-MS

Xhixha, G., Bezzon, G.P., Broggini, C., Buso, G.P., Caciolli, A., Callegari, I., De Bianchi, S., Fiorentini, G., Guastaldi, E., Kaçeli Xhixha, M., Mantovani, F., Massa, G., Menegazzo, R., Mou, L., Pasquini, A., Rossi Alvarez, C., Shyti, M., 2013. The worldwide NORM production and a fully automated gamma-ray spectrometer for their characterization. *Journal of Radioanalytical and Nuclear Chemistry*. 295, 445–457. Doi: 10.1007/s10967-012-1791-1